\documentclass[conference]{IEEEtran}
\usepackage{amsmath,amssymb,amsfonts}
\usepackage{algorithmic}
\usepackage{graphicx}
\usepackage{textcomp}
\usepackage{xcolor}
\def\BibTeX{{\rm B\kern-.05em{\sc i\kern-.025em b}\kern-.08em
    T\kern-.1667em\lower.7ex\hbox{E}\kern-.125emX}}

\usepackage[square,numbers]{natbib}
\usepackage{balance}
\usepackage{lipsum}
\usepackage{tcolorbox}
\usepackage[inline]{enumitem}
\usepackage{enumerate}
\usepackage{multirow}
\usepackage{subcaption}
\usepackage[scientific-notation=true]{siunitx}
\usepackage{hyperref} 

\begin{document}




\title{Quantum vs. Classical Machine Learning Algorithms for Software Defect Prediction: Challenges and Opportunities}






\author{
    \IEEEauthorblockN{
    Md Nadim\IEEEauthorrefmark{1}, 
    Mohammad Hassan\IEEEauthorrefmark{2},
    Ashis Kumar Mandal\IEEEauthorrefmark{1},   
    Chanchal K. Roy\IEEEauthorrefmark{1}
    }
    \IEEEauthorblockA{
        \IEEEauthorrefmark{1}Department of Computer Science, University of Saskatchewan, Saskatoon, Canada \\
        \IEEEauthorrefmark{2}University of Prince Edward Island, Charlottetown, PE, Canada\\
        Email: \{mnadims.cse, ashis.62, mmmhbd76\}@gmail.com,
        \IEEEauthorrefmark{2}chanchal.roy@usask.ca
    }
}

\maketitle

\begin{abstract}

Software defect prediction is a critical aspect of software quality assurance, as it enables early identification and mitigation of defects, thereby reducing the cost and impact of software failures. Over the past few years, quantum computing has risen as an exciting technology capable of transforming multiple domains; Quantum Machine Learning (QML) is one of them. QML algorithms harness the power of quantum computing to solve complex problems with better efficiency and effectiveness than their classical counterparts. However, research into its application in software engineering to predict software defects still needs to be explored. In this study, we worked to fill the research gap by comparing the performance of three QML and five classical machine learning (CML) algorithms on the 20 software defect datasets. Our investigation reports the comparative scenarios of QML vs. CML algorithms and identifies the better-performing and consistent algorithms to predict software defects. We also highlight the challenges and future directions of employing QML algorithms in real software defect datasets based on the experience we faced while performing this investigation. The findings of this study can help practitioners and researchers further progress in this research domain by making software systems reliable and bug-free.
\end{abstract}





\section{Introduction}
\label{introduction}
Software quality assurance \cite{2019:Borg:SZZUnleashed} is an enduring concern in the realm of software development, with a principal goal of minimizing defects in the software system \cite{Rel:2015:Yang:Deep:JIT:Defect:Prediction}. A software defect (also known as a bug) can be an error, flaw, failure, or fault in a computer program or system that prevents a software system from working as desired or produces an incorrect or unexpected result. Defects in software projects can undermine its reliability and efficiency. Several studies \cite{Nadim02, Nadim03} propose different techniques to detect and predict these software bugs to ensure reliable and efficient software systems. The early detection \cite{Kim:Classify:Clean:Buggy} and effective mitigation \cite{APR-Patch-Code} of software defects are paramount, as they can lead to substantial costs and detrimental consequences \cite{Has-bug-really-fixed}.  In pursuit of this objective, the software development community has continuously sought innovative technologies and methodologies to advance the effectiveness of defect prediction techniques.  

In recent years, quantum computing \cite{QuantumComputing} has emerged as a pivotal area of interest across various scientific and technological domains. With its distinctive ability to manipulate information using quantum bits or qubits, quantum computing promises to address complex problems more effectively than classical computing paradigms. One particularly promising avenue is the fusion of quantum computing with machine learning, a field known as Quantum Machine Learning (QML), which is a relatively new domain that combines ideas from Machine Learning (ML), Quantum Computing (QC), and Quantum Information (QI) \cite{Pushpak2021}.   

Our study conducts a comprehensive performance evaluation of three QML algorithms, and we compare their effectiveness with five widely used classical machine learning (CML) methods commonly employed for software defect prediction. These machine learning algorithms are readily accessible through the Qiskit library \cite{Qiskit} within the Python programming language, and they encompass the following:
Pegasos Quantum Support Vector Classifier (PQSVC),
Quantum Support Vector Classifier (QSVC), and
Variational Quantum Classifier (VQC). We employ a set of well-established CML classifiers conveniently accessible through the Scikit-learn library  \cite{scikit-learn} within the Python programming environment: Support Vector Classifier (SVC), Random Forest (RF) Classifier, K-Nearest Neighbors (KNN) Classifier, Gradient Boosting Classifier (GBC), and Perceptron (PCT). 

To rigorously assess and compare the effectiveness of these methods, we employ a set of critical metrics, including the number of accurately predicted buggy and clean commits, precision, recall, F1 score, and run time for the ML algorithms. The assessment results of this study are not limited to performance evaluations only. They aim to advance our understanding of the possibilities and limitations of quantum computing in the software defect prediction domain. We do not find any prior study to evaluate such performance comparison of Quantum and Classical ML algorithm in software defect prediction datasets. Moreover, this research offers meaningful insights for practitioners and researchers who wish to use QML algorithms to improve software quality assurance processes.  

We evaluated our results to answer the following research questions (\textbf{RQs}).

\noindent 
\textbf{RQ1: What are the primary impediments to using the QML algorithms for software defect datasets?}
  
\begin{itemize} 
\item We are \textbf{compelled} to address this research question (RQ) due to the emergence of QML algorithms as a promising approach for various practical applications in ML techniques. However, their application in defect predictions within software systems remains significantly underexplored. By delving into this RQ, we aim to uncover the practical implementation challenges, especially concerning scalability, interpretability, and resource requirements. By identifying and understanding these challenges, researchers can devise appropriate strategies to adapt and implement necessary changes in real-world scenarios. 
\item 
We \textbf{delve} into this RQ, recognizing the current stage of development of real quantum computers, which remain largely inaccessible. As a result, the implementation of QML algorithms requires the utilization of quantum simulators. Throughout our study, we have thoroughly scrutinized the obstacles encountered when working with a diverse array of software defect datasets, spanning different sizes and application domains. The insights collected from our investigation offer valuable guidance for answering this RQ.
\end{itemize} 

\noindent
\textbf{RQ2: How effective is QML over CML for software bug prediction across different project domains?}

\begin{itemize}
\item Like the \textbf{rationale} behind investigating the robustness of different QML algorithms, comparing the effectiveness of QML over CML for software bug prediction across different project domains lies in the importance of understanding which machine learning approach yields better results in this specific context. With this comparison we can gain insights into their respective strengths and weaknesses, helping developers and researchers make informed decisions about which approach to use in different scenarios. Additionally, examining their performance across various project domains allows for a more comprehensive understanding of their generalizability and applicability in real-world settings.
\item To \textbf{investigate} this research question, we conducted a comprehensive comparative analysis involving three QML and five CML algorithms. We evaluated their performance across 14 subject systems containing 20 datasets from diverse domains. Our evaluation encompassed multiple aspects, comparing QML algorithms to other QML algorithms, QML algorithms to CML algorithms, and CML algorithms to one another. We assessed their performance based on three key metrics, including the number of correctly predicted buggy and clean commits, precision, recall, F1 score, and the time required to execute these ML algorithms on each dataset.
\end{itemize}

To enhance the reproducibility of our research findings, we have made the complete replication package accessible online \cite{figshareNadim2024}. The replication package includes all necessary commands to install the Python environment, datasets, and code to replicate the experiments conducted in this study. 

In this paper, the sections are organized as follows: Section \ref{background} covers the study’s background, while Section \ref{methodology} details our methodology. Section \ref{perform-compare} presents the performance comparison, discusses the findings, and addresses the study’s research questions. Section \ref{threats-validity} identifies potential threats to validity, and Section \ref{related-work} reviews related literature. Finally, Section \ref{conclusion-future} concludes the paper and highlights future research directions.

\section{Background}
\label{background}
\textbf{Quantum computing} is a new approach to computing in which the principles of quantum mechanics are harnessed to solve complex problems that are computationally expensive for classical computers \cite{hidary2019quantum}. The fundamental unit of quantum computing is the qubit, and it is a linear combination of the zero and one states. In other words, a qubit can exist in a superposition of these two states, whereas a classical computer can only be in one state at a time. With the advantage of quantum mechanical effects such as superposition, entanglement, and interference, quantum computers can solve complex problems \cite{khrennikov2021roots}. 

Different \textbf{types of quantum computers} have been developed, with the gate model and the quantum annealing model being the most prominent \cite{ORUS2019100028}. In quantum computers, qubits are manipulated using various quantum gates to solve problems in the quantum gate model. In contrast, the quantum annealing model \cite{Dwave} utilizes the adiabatic theorem of quantum mechanics to solve the problem. 


\textbf{Quantum Machine Learning (QML)} is an emerging field that leverages quantum information processing to enhance machine learning tasks, including clustering, regression, and classification. By utilizing quantum operations, QML can efficiently manage high-dimensional data and uncover complex patterns beyond the reach of many classical machine learning (CML) algorithms \cite{ramezani2020machine}. QML approaches are categorized based on the nature of both the data (classical or quantum) and the machine learning algorithms employed (classical or quantum), resulting in four distinct classifications \cite{Alchieri2021}. 

A \textbf{Variational Quantum Classifier (VQC)} \cite{9665779} is a classical hybrid QML algorithm that uses a PQC for data encoding and processing, facilitating classification tasks. In the training phase, the parameters of quantum circuits are fine-tuned using classical optimization techniques to reduce classification errors. During prediction, testing data is converted into a quantum state, and the VQC makes class-label predictions based on quantum state characteristics.

A \textbf{Quantum Support Vector Classifier (QSVC)} is a QML algorithm designed for binary classification based on Support Vector Machines (SVM). To date, several QSVC algorithms have been proposed \cite{ramezani2020machine}. The primary distinction between quantum and classical SVC lies in the fact that, in the classical case, the explicit kernel function is often known, whereas in quantum computing, it is calculated using a quantum circuit \cite{Heredge2021}. In QSVC, input data is encoded into a quantum state mapped to a high-dimensional quantum feature space, and quantum kernels aid in maximizing the margin between classes while minimizing classification errors.

\section{Methodology}
\label{methodology}
\begin{figure}
\centering
\includegraphics[width=0.45\textwidth] {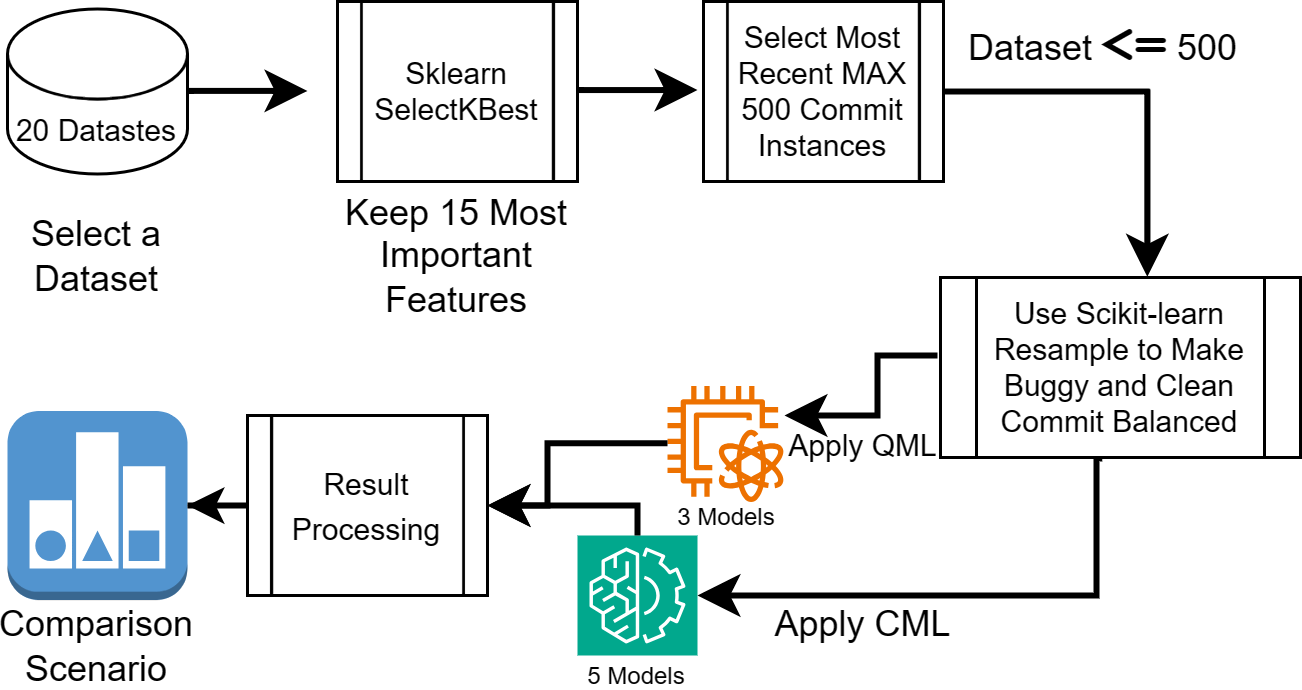}
\caption{Overall Methodology}
\label{fig:methodology}
\end{figure}

In this section, we describe the comprehensive methodology used to leverage three QML algorithms and five CML algorithms across a curated set of 20 datasets. We provide a visual representation of the distinct stages of our investigation in Figure \ref{fig:methodology}, which is iteratively applied to every dataset within our collection. Our methodology comprises the following key stages.

\begin{table}[]
\centering
\caption{Manually Labeled Dataset \cite{Nadim03}}
\label{tab:data-summary-i}
\begin{tabular}{|llc|c|}
\hline
\multicolumn{1}{|l|}{\textbf{\begin{tabular}[c]{@{}l@{}}Subject Systems\end{tabular}}} & \multicolumn{1}{l|}{\textbf{Application Domains}}                                                               & \textbf{\begin{tabular}[c]{@{}c@{}}Prg.\\ Lang.\end{tabular}} & \textbf{\begin{tabular}[c]{@{}c@{}}Data \\ Instances\end{tabular}} \\ \hline \hline
\multicolumn{1}{|l|}{Accumulo}                                                              & \multicolumn{1}{l|}{\begin{tabular}[c]{@{}l@{}}Distributed   NoSQL\\ database for big data\end{tabular}}       & Java                                                           & 102                                                                \\ \hline
\multicolumn{1}{|l|}{Ambari}                                                                & \multicolumn{1}{l|}{\begin{tabular}[c]{@{}l@{}}Cluster   management and\\ monitoring tool\end{tabular}}       & Java                                                           & 68                                                                 \\ \hline
\multicolumn{1}{|l|}{Hadoop}                                                                & \multicolumn{1}{l|}{\begin{tabular}[c]{@{}l@{}}Distributed   big \\ data framework\end{tabular}}               & Java                                                           & 102                                                                \\ \hline
\multicolumn{1}{|l|}{Jackrabbit}                                                            & \multicolumn{1}{l|}{\begin{tabular}[c]{@{}l@{}}Content   repository \\ for Java applications\end{tabular}}     & Java                                                           & 116                                                                \\ \hline
\multicolumn{1}{|l|}{Lucene}                                                                & \multicolumn{1}{l|}{\begin{tabular}[c]{@{}l@{}}Text   Search \\ Engine Library\end{tabular}}                   & Java                                                           & 262                                                                \\ \hline
\multicolumn{1}{|l|}{Oozie}                                                                 & \multicolumn{1}{l|}{\begin{tabular}[c]{@{}l@{}}Runs   Hadoop Workloads\\ through Web Services\end{tabular}} & Java                                                           & 92                                                                 \\ \hline \hline
\multicolumn{3}{|r|}{\textbf{Total}}                                                                                                                                                                                                                                          & \textbf{742}                                                       \\ \hline
\end{tabular}
\end{table}
\begin{table}
\centering
\caption{Automatically Labeled Dataset \cite{Nadim02, Nadim03}}
\label{tab:data-summary-ii}
\begin{tabular}{|llc|c|}
\hline
\multicolumn{1}{|l|}{\textbf{\begin{tabular}[c]{@{}l@{}}Subject Systems\end{tabular}}} & \multicolumn{1}{l|}{\textbf{Application Domains}}                                                               & \textbf{\begin{tabular}[c]{@{}c@{}}Prg.\\ Lang.\end{tabular}} & \textbf{\begin{tabular}[c]{@{}c@{}}Data \\ Instances\end{tabular}} \\ \hline \hline
\multicolumn{1}{|l|}{Accumulo}                                                             & \multicolumn{1}{l|}{\begin{tabular}[c]{@{}l@{}}Distributed NoSQL \\ database for big   data\end{tabular}}       & Java                                                          & 200                                                                \\ \hline
\multicolumn{1}{|l|}{Ambari}                                                               & \multicolumn{1}{l|}{\begin{tabular}[c]{@{}l@{}}Cluster management \\ and monitoring   tool\end{tabular}}        & Java                                                          & 282                                                                \\ \hline
\multicolumn{1}{|l|}{Bitcoin}                                                              & \multicolumn{1}{l|}{\begin{tabular}[c]{@{}l@{}}Cryptocurrency based \\ on blockchain   technology\end{tabular}} & CPP                                                           & 412                                                                \\ \hline
\multicolumn{1}{|l|}{Camel}                                                                & \multicolumn{1}{l|}{\begin{tabular}[c]{@{}l@{}}Integration framework \\ for routing   data\end{tabular}}        & Java                                                          & 682                                                                \\ \hline
\multicolumn{1}{|l|}{Jackrabbit}                                                           & \multicolumn{1}{l|}{\begin{tabular}[c]{@{}l@{}}Content repository for \\ Java   applications\end{tabular}}      & Java                                                          & 456                                                                \\ \hline
\multicolumn{1}{|l|}{Jenkins}                                                              & \multicolumn{1}{l|}{\begin{tabular}[c]{@{}l@{}}Automation server for \\ software   development\end{tabular}}    & Java                                                          & 314                                                                \\ \hline
\multicolumn{1}{|l|}{Litecoin}                                                             & \multicolumn{1}{l|}{\begin{tabular}[c]{@{}l@{}}Cryptocurrency based \\ on blockchain   technology\end{tabular}} & CPP                                                           & 394                                                                \\ \hline
\multicolumn{1}{|l|}{Lucene}                                                               & \multicolumn{1}{l|}{\begin{tabular}[c]{@{}l@{}}Open-source search\\ library in   Java\end{tabular}}             & Java                                                          & 206                                                                \\ \hline
\multicolumn{1}{|l|}{Mongo}                                                                & \multicolumn{1}{l|}{\begin{tabular}[c]{@{}l@{}}NoSQL database \\ system\end{tabular}}                           & CPP                                                           & 360                                                                \\ \hline
\multicolumn{1}{|l|}{Oozie}                                                                & \multicolumn{1}{l|}{\begin{tabular}[c]{@{}l@{}}Workflow scheduler \\ for Hadoop   ecosystem\end{tabular}}       & Java                                                          & 310                                                                \\ \hline
\multicolumn{1}{|l|}{OpenStack}                                                             & \multicolumn{1}{l|}{\begin{tabular}[c]{@{}l@{}}Open-source cloud\\  computing platform\end{tabular}}   & Python                                                        & 270                     \\ \hline 
\multicolumn{1}{|l|}{QT}                                                                    & \multicolumn{1}{l|}{\begin{tabular}[c]{@{}l@{}}Cross-platform  \\  C++ framework \\ and tools\end{tabular}} & CPP                                                           & 658                     \\ \hline
\multicolumn{1}{|l|}{Tomcat}                                                               & \multicolumn{1}{l|}{\begin{tabular}[c]{@{}l@{}}Servlet container for \\ Java   applications.\end{tabular}}      & Java                                                          & 324                                                                \\ \hline
\multicolumn{1}{|l|}{Zxing}                                                                & \multicolumn{1}{l|}{\begin{tabular}[c]{@{}l@{}}Open-source barcode \\ image   processing\end{tabular}}          & Java                                                          & 354                                                                \\ \hline \hline
\multicolumn{3}{|r|}{\textbf{Total}}                                                                                                                                                                                                                                         & \textbf{5,222}                                                      \\ \hline
\end{tabular}
\end{table}

\subsection{Selecting the Datasets}
We initially acquired the 20 datasets to be used in this study from two recent studies \cite{Nadim02, Nadim03} for the enhancement of buggy commit detection from software systems. Each of these datasets sheds light on defects encountered in different revisions of software projects and previously employed to enhance the efficacy of identifying defects within software commits utilizing CML and Deep Learning approaches. We have specifically chosen these datasets to comprehensively evaluate the practical challenges and the comparative performance of QML algorithms versus Classical Machine Learning (ML) algorithms. 

These datasets contain a large number of columns (features), which require to be reduced to use them using QML algorithms. QML algorithms can not work for a large number of features because of the complexities in converting these features into the quantum feature map. Different subject systems of collected datasets \cite{Nadim02, Nadim03} contain a range from 40 to thousands of data columns extracted from n-gram-based source code token sequence (TS) and token patterns (TP) with conventional features used in software bug prediction studies. Tokens are considered as the smallest element of any programming language, such as a variable name, a keyword, an operator, etc. To prepare these datasets for QML and CML processing, we first employ Pearson Correlation \cite{PearsonCorrelation} and bring all these datasets into 32 columns/ features by eliminating less correlated data columns to the target label. Then, we use Scikit-learn \cite{scikit-learn} K-best feature selection methods to exclude more irrelevant data columns to bring the feature count within 15. This initial step serves the dual purpose of reducing dataset dimensionality and enhancing the performance of both QML and CML algorithms by eliminating the influence of irrelevant feature values.

Additionally, the number of features in each dataset corresponds to its dimensionality, a critical factor. Reducing dimensionality not only aids model performance but also influences the number of qubits necessary for QML algorithms. We initially attempted to run QML algorithms with higher dimensionality, but in most cases, this resulted in failure for datasets with dimensionality exceeding 15 with qubits more than 8. Following dimensionality reduction and minimizing the number of qubits to 8, we apply the processed datasets to all QML and CML algorithms.

Among the 20 datasets examined, a subgroup of six datasets in Table \ref{tab:data-summary-i} includes software commits with known labels that have undergone manual classification as either buggy or clean. This specific subset of the dataset serves by offering valuable insights into the challenges of human-guided defect labeling. As manually labeling software bug to each of its commits is a slow and time-consuming process, the data subset in this table have lower number of data instances.  In addition to this data subset, we also investigate 14 distinct subject systems listed in Table \ref{tab:data-summary-ii}, encompassing automatically labeled data from a wide array of software projects having diverse application domains. The datasets incorporate three distinct categories of feature values, namely GitHub Statistics (GS), Source Code Graph (SCG), and Token Sequence and Pattern (TS-TP) based features, as previously outlined in earlier research studies \cite{Nadim02, Nadim03}.

The software projects we investigate in this study have been developed using a variety of programming languages, including CPP, Java, and Python. This diversity mirrors the real-world programming landscape, enhancing the practicality and relevance of our research findings. The provided data tables offer valuable insight into the distribution of buggy and clean commits in various software projects. These datasets also encompass a wide range of subject systems from diverse application domains, such as Data Storage and Retrieval, System Administration, Distributed Computing, Content Repository, Text Search, and more. The total commit instances across all datasets, amounting to 5,964, demonstrate the extensive scale of the study. Such a scale is crucial for drawing statistically significant conclusions and generalizing findings, offering a robust foundation for research and analysis. As such, a comparative investigation of quantum and classical ML algorithms has yet to be available. Our study can shed light on a diverse comparative scenario of the effectiveness of QML vs. CML algorithms and inspire more future studies in similar domains.

\subsection{Selecting the Optimal Feature Sets}
Machine learning algorithms are highly impacted by the number of attributes or features in the data they operate on. The datasets we worked with in our study contained a substantial number of features, more than 100 in each case. Initially, we attempted to apply QML algorithms with a larger number of features, but we encountered challenges. Most QML algorithms failed to produce meaningful results when dealing with datasets containing more than 15 features.

We took a stepwise approach to eliminate the irrelevant features from the datasets and make them manageable by the QML algorithms. We first employed the Pearson correlation \cite{PearsonCorrelation} method to identify the best 32 features that exhibited a strong positive correlation with the target output variable. We apply the Pearson correlation method as it is widely used in research due to its simplicity and ease of interpretation, making it applicable to various data types and fields. Subsequently, we used the SelectKBest algorithm from the Scikit-Learn library \cite{scikit-learn} in the Python programming language to narrow the selection to the top 15 features. These 15 features were used in QML and CML algorithms. This process allowed us to work with a manageable and meaningful subset of features for our analysis.

\subsection{Applying QML and CML Algorithms}
In our research, we take a systematic approach to comparing quantum and classical ML algorithms. First, we carefully choose the datasets and decide which features to use. Once that is done, we use three quantum QML and five CML algorithms. We use 70\% and 30\% data instances to train and test each algorithm, respectively. To ensure a fair comparison, we make sure that all the algorithms have similar parameter settings (e.g., selection of random seed, feature optimizer, number of qubits, quantum kernel, etc.) algorithms,  to allow us to make a meaningful and balanced evaluation of quantum and CML algorithms.

\subsection{Dataset Size Vs. Execution Time}
\label{exeTime}
\begin{figure}
\centering
\includegraphics[width=0.48\textwidth] {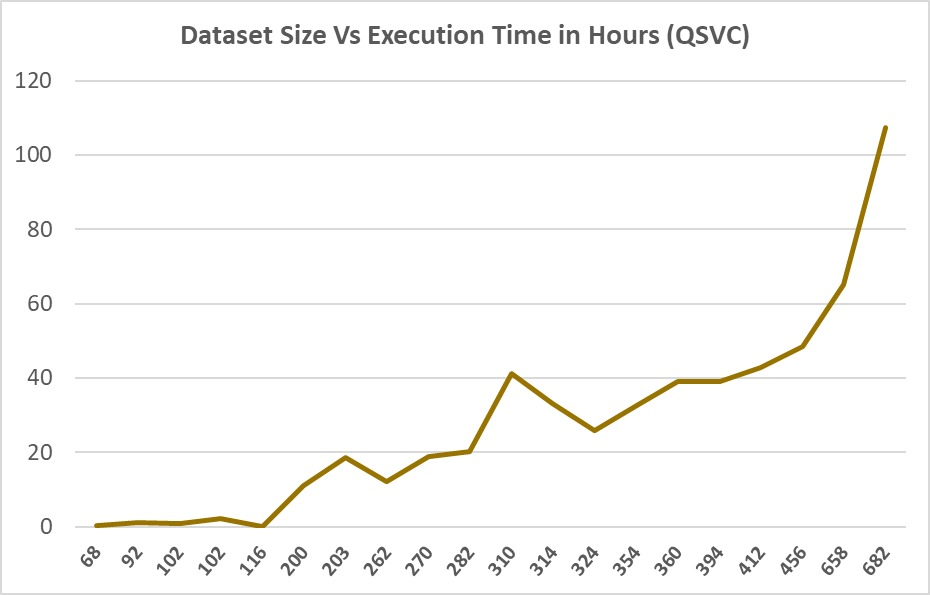}
\caption{Comparing Run Time on Different Dataset Size using QSVC Algorithm}
\label{fig:averageRunTime}
\end{figure}

Execution time is very crucial while applying QML algorithms as it takes a longer time compared to the Classical approaches. Our preliminary investigation of all the QML and CML algorithms showed that QSVC takes much longer time compared to all other QML and CML algorithms. Figure \ref{fig:averageRunTime} shows the time taken in hours by the QSVC algorithm to get the output from all the 20 datasets used in this investigation. The X-axis of the diagram represents the number of dataset instances to be processed from each datasets, mentioned in Tables \ref{tab:data-summary-i} \& \ref{tab:data-summary-ii}. The Y-axis of the diagram represents the time (in hours) taken to produce the output for each dataset by the QSVC algorithm. 

The investigation into the relationship between dataset size and execution time for the QSVC algorithm reveals a noteworthy trend. As the dataset approaches a threshold of 400 to 600, the execution time increases too rapidly and surpasses 100 hours (more than 4 days), highlighting a critical point in scalability. The observed pattern in the line graph suggests that the growth of execution time is indicative of an exponential nature. This exponential increase in execution time raises concerns about the responsiveness of the QSVC algorithm as dataset size expands. 

Consequently, to comprehensively assess and compare the performance of various Quantum and Classical Machine Learning (QML and CML) algorithms, we deliberately constrained the dataset size to get output for all the QML and CML algorithms. This is the reason we truncated the data instances, keeping them within 700. This strategic approach allows for a more nuanced understanding of algorithmic behaviors and performance under controlled conditions, ensuring a comprehensive evaluation of their capabilities.

To achieve this, we conducted experiments on a high-performance Linux Server equipped with a 64-core Intel(R) Xeon(R) Platinum 8356H CPU running at 3.90GHz, with a CPU MHz of 4399.999, a generous cache size of 36608 KB, and a whopping total memory of 3.14 terabytes. We only presented the results of the QSVC algorithm in Figure \ref{fig:averageRunTime}, as the run times for all other algorithms were significantly shorter in comparison. Our investigation focused on capturing the runtime of the QSVC algorithm alongside the cut-off threshold to ensure a comprehensive understanding of its efficiency and performance characteristics in this investigation.

Although QML algorithms currently exhibit slower runtimes, we believe that, as promised, this limitation will diminish once real quantum computers become widely available and efficient. For now, we can focus on exploring use-case scenarios to ensure the algorithms are properly leveraged when quantum computers are more accessible.


\section{Performance Comparison \& Discussion}
\label{perform-compare}
Our in-depth analysis of the experimental results uncovers valuable insights into the performance of the QML and CML algorithms used in this study. We demonstrate the insights from our findings in Figure \ref{fig:averageRunTime} \& Table \ref{tab:fscoreQMLCML}. In this section, we compare performance of different QML and CML algorithms and answer the research questions (RQs) based on these findings.  


\subsection{Challenges in Applying QML}
\label{RQ3challenges}
Recent studies \cite{Cerezo2022} have begun to delve into the complex challenges associated with the application of QML methodologies in various research domains, including biochemistry and high-energy physics. 
Our focus shifts to the fascinating realm of software defect prediction, a crucial domain in the software development landscape. In the following, we describe the unique challenges encountered when applying QML in the context of our study. 


The \textbf{availability of Real Quantum Computers} remains limited. While IBM Q \cite{IBMQ}, Rigetti \cite{RigettiQ}, and IonQ \cite{IonQ} offer cloud services for conducting experiments on their Quantum Computers, it still presents a significant barrier to access for many researchers in the field due to underlying cost and availability. This scarcity compels researchers to rely on quantum simulators, which are notably slower and pose unique challenges when confronted with real-world software engineering scenarios. 



\begin{tcolorbox}[colback=gray!10!white, colframe=gray!80!black, arc=4mm]
\textbf{RQ1: What are the primary impediments to using the QML algorithms for software defect datasets?}
Key challenges include the limited availability of real quantum computers, inefficient quantum bits (qubits), handling large datasets with a large number of features (dimensions), and improving the interpretability of quantum models.
\end{tcolorbox}

\textbf{Limited Quantum Bits (Qubits)} pose a significant challenge when applying QML algorithms to real-world datasets. Initially, these algorithms map classical datasets to their quantum representations through a process known as quantum feature mapping. The complexity of this mapping process grows exponentially with the number of qubits and the features in the dataset. Consequently, a lower number of qubits and features results in faster processing of QML algorithms. 

In our experiments, we encountered processing bottlenecks when dealing with datasets comprising 30 feature values using 30 qubits in our preliminary trials. To address this, we iteratively reduced both the number of feature values and qubits in smaller datasets, aiming to identify an optimal balance that could deliver results within a reasonable timeframe. We employ Scikit-learn \cite{scikit-learn} K-best feature selection methods to extract the 15 most important features from the datasets, a strategic approach for conducting our experiments. This feature reduction is pivotal, as handling an extensive feature set without mitigation becomes highly challenging for QML algorithms and often leads to the failure of most QML algorithms to produce meaningful results. Our investigation concluded when we reached a configuration of (8 qubits, 15 feature values), which successfully produced outputs for all subject systems. However, even with this optimized configuration, the processing time for the largest dataset (Camel in Table \ref{tab:data-summary-ii}), consisting of 682 data instances, exceeded 100 hours (\ref{fig:averageRunTime}). This underscores the computational challenges inherent in scaling QML algorithms to handle real-world datasets effectively.

\textbf{Quantum Feature Mapping} presents a unique challenge, requiring the transformation of classical data into a quantum state. The selection of effective quantum feature maps is very important for the success of quantum machine learning applications. In our approach, we utilize the power of the ZZFeatureMap from the IBM Qiskit library \cite{Qiskit}. Within the IBM Qiskit's Circuit Library \cite{IBMQ}, three distinct feature mapping techniques are available: PauliFeatureMap, ZFeatureMap, and ZZFeatureMap. 
The ZZFeatureMap introduces qubit entanglement, a feature absent in the ZFeatureMap, making it the preferred choice for this investigation. It is important to note that the selection of quantum machine learning algorithms can influence the choice of suitable feature mapping methods. In our future research, we plan to explore the relationship between different quantum feature mapping techniques, specifically in the context of software bug detection datasets.

The \textbf{Interpretability of QML Algorithms} often exhibits complex and intricate internal representations, making them less interpretable than CML algorithms. This lack of interpretability can hinder the identification of the root causes of their performance, as it is challenging to understand why a particular prediction was made. Tasks to make QML algorithms more transparent and interpretable are yet to be done on a large scale. Once this challenge is addressed, it can resolve this challenge in the future. 

\begin{figure*}
\centering
\includegraphics[width=0.92\textwidth] {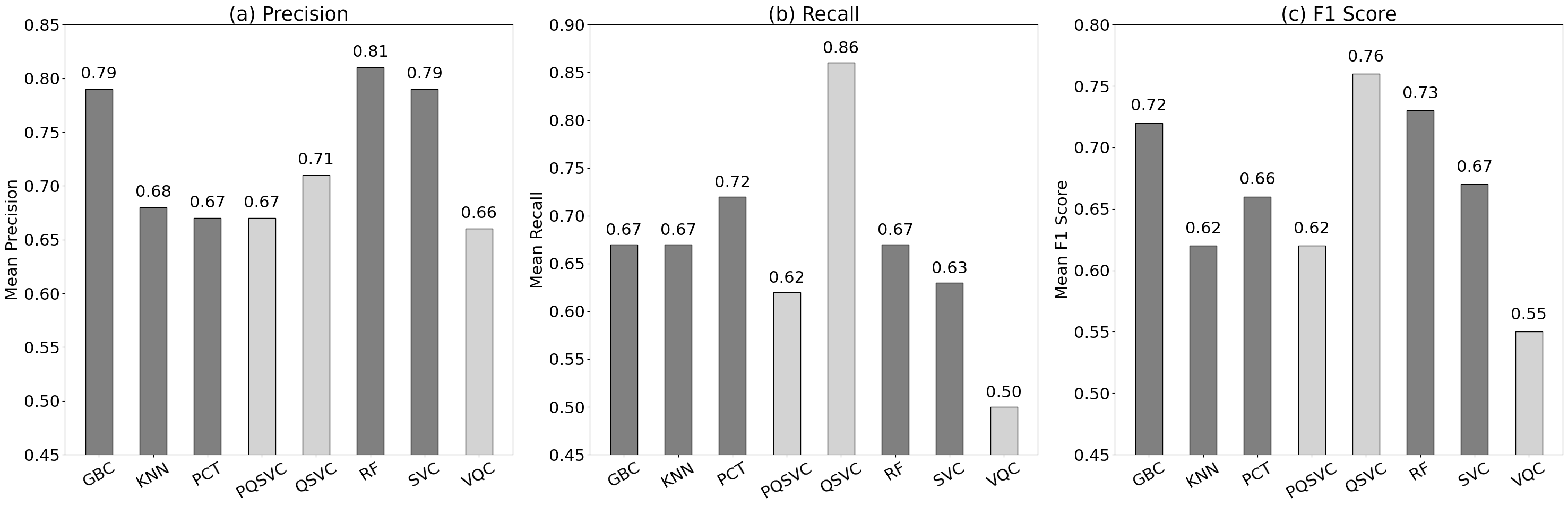}
\caption{Average Precision, Recall, and F1 Scores for CML and QML algorithms are illustrated in this figure. CML algorithms are depicted in dark colors, while QML algorithms are represented in lighter shades}
\label{fig:meanMetric}
\end{figure*}

\begin{table}[ht]
\centering
\caption{Comparing F1 Score Among QML and Better Performing CML Algorithms On 20 Datasets}
\label{tab:fscoreQMLCML}
\begin{tabular}{l|ccc|ccc}
\hline
\textbf{Dataset} & \multicolumn{3}{c|}{\textbf{QML Algorithms}} & \multicolumn{3}{c}{\textbf{CML Algorithms}} \\ \hline
                  & \textbf{PQSVC} & \textbf{QSVC} & \textbf{VQC} & \textbf{GBC} & \textbf{RF} & \textbf{SVC} \\ \hline
1                 & 0.45           & \textbf{0.78}          & 0.33         & 0.58         & 0.65        & 0.75         \\
2                 & 0.72           & \textbf{0.83}          & 0.56         & 0.74         & 0.76        & 0.68         \\
3                 & 0.53           & 0.76          & 0.61         & 0.78         & \textbf{0.86}        & 0.77         \\
4                 & 0.64           & 0.85          & 0.62         & 0.87         & \textbf{0.88}        & 0.82         \\
5                 & 0.80           & \textbf{0.85}          & 0.72         & 0.78         & 0.82        & 0.76         \\
6                 & 0.49           & \textbf{0.78}          & 0.49         & 0.71         & 0.61        & 0.73         \\
7                 & 0.63           & 0.76          & 0.68         & 0.77         & \textbf{0.82}        & 0.77         \\
8                 & 0.69           & \textbf{0.72}          & 0.57         & 0.67         & 0.63        & 0.61         \\
9                 & 0.67           & 0.58          & 0.69         & \textbf{0.78}         & 0.75        & 0.76         \\
10                & 0.64           & 0.73          & 0.36         & \textbf{0.77}         & \textbf{0.77}        & 0.75         \\
11                & 0.55           & \textbf{0.71}          & 0.60         & 0.61         & 0.70        & 0.70         \\
12                & 0.72           & \textbf{0.75}          & 0.55         & 0.68         & 0.67        & 0.72         \\
13                & 0.50           & \textbf{0.86}          & 0.62         & 0.76         & 0.80        & 0.49         \\
14                & 0.71           & \textbf{0.80}          & 0.64         & \textbf{0.80}         & 0.76        & 0.67         \\
15                & \textbf{0.74}           & \textbf{0.74}          & 0.31         & 0.59         & 0.59        & 0.62         \\
16                & 0.52           & \textbf{0.78}          & 0.36         & 0.69         & 0.69        & 0.69         \\
17                & \textbf{0.62}           & \textbf{0.62}          & \textbf{0.62}         & \textbf{0.62}         & \textbf{0.62}        & \textbf{0.62}         \\
18                & 0.50           & \textbf{0.88}          & 0.60         & 0.75         & 0.72        & 0.52         \\
19                & 0.70           & 0.71          & 0.56         & \textbf{0.77}         & 0.70        & 0.44         \\
20                & 0.56           & 0.68          & 0.51         & 0.72         & \textbf{0.73}        & 0.61         \\ \hline
\textbf{AVG}      & 0.62  & \textbf{0.76} & 0.55 & 0.72 & 0.73 & 0.67 \\ \hline
\end{tabular}
\end{table}

\subsection{Comparing QML vs. CML Algorithms}
\label{QMLvsCML}
In this study, we performed a comparative analysis of three QML algorithms — PQSVC, QSVC, and VQC—and five Classical Machine Learning (CML) algorithms — Support Vector Classifier (SVC), Random Forest (RF), K-Nearest Neighbors (KNN), Gradient Boosting Classifier (GBC), and Perceptron (PCT). These algorithms were evaluated across 20 datasets to predict whether a given software commit is \textbf{buggy} or \textbf{clean}. The F1 scores and their averages for the different QML and top-performing CML algorithms are presented in Table \ref{tab:fscoreQMLCML}. 

Figure \ref{fig:meanMetric} shows the average precision, recall, and F1 scores for predicting buggy software instances across all 20 datasets, using a variety of QML and CML algorithms. The bars representing CML algorithms are depicted in dark gray, while those for QML algorithms are shown in light gray. This visual comparison highlights the performance of QML algorithms, including PQSVC, QSVC, and VQC, against five CML algorithms: GBC, KNN, PCT, RF, and SVC. These metrics provide a comparative evaluation of each algorithm's ability to correctly identify buggy instances in the datasets. Among these results, we can see that the QML algorithm (QSVC) is performing better in Recall and F1~Score compared to the other two QML and five CML algorithms. These results highlight QSVC's superior ability to accurately classify software commits compared to the other QML algorithms. Considering Precision, CML algorithms, GBC, RF, and SVC, outperform all the other QML and CML algorithms.

\begin{tcolorbox}[colback=gray!10!white, colframe=gray!80!black, arc=4mm]
\textbf{RQ2: How effective is QML over CML for software bug prediction across different project domains?}

The comparison in Table \ref{tab:fscoreQMLCML} reveals that the QSVC algorithm outperforms other QML and CML algorithms in predicting buggy software instances, achieving the highest F1 score in 13 out of 20 datasets. In contrast, the RF, GBC, PQSVC, VQC, and SVC algorithms lead in 6, 5, 2, 1, and 1 datasets, respectively, highlighting their superior performance detecting buggy commits. This analysis demonstrates the varying strengths of QML and CML algorithms in different datasets for bug prediction tasks.
\end{tcolorbox}

Overall, when considering the individual Recall and F1 scores of QML algorithms and their average performance across all datasets, QSVC stands out as the best-performing algorithm, surpassing both PQSVC and VQC. While considering precision, inferior performance of QML algorithms depicts more number of false positives compared to the CML algorithms. Although, the performance of QML algorithms are not superior in Precision, it is comparable with two CML algorithms, KNN and PCT. This comparative analysis concludes that, although QML algorithms may not always offer superior Precision, their better Recall and F1 scores make them a viable choice for software defect prediction, depending on the desired balance between Recall and Precision.

\section{Threats to Validity}
\label{threats-validity}
\textbf{Internal Validity:} The primary threat to the internal validity of our study arises from the use of quantum simulators rather than real quantum hardware. Quantum simulators are approximations of quantum computing systems, and while they aim to mimic quantum behavior, they may not fully capture the complexities or imperfections of actual quantum devices. This discrepancy can potentially introduce biases in the performance outcomes of QML algorithms. To mitigate this, we evaluated three QML algorithms across 20 datasets, using three distinct performance metrics, resulting in a total of 180 performance values. This robust evaluation was designed to account for any inconsistencies inherent in the simulator environment. However, there remains a possibility that the simulator could present conditions that are not reflective of real quantum systems. Although our comprehensive analysis reduces the potential impact of these simulator-induced limitations, further validation on actual quantum hardware is necessary to confirm the generalizability of our findings.

\textbf{External Validity:} The external validity of our study is limited by the use of a specific set of QML algorithms implemented in the Qiskit library. While Qiskit is a widely recognized and accessible quantum computing library, other quantum machine learning libraries, such as PennyLane \cite{PennyLane}, TensorFlow Quantum \cite{broughton2021tensorflow}, and Yao \cite{Luo2020YaoQuantum}, offer different sets of algorithms. Use of any of these systems could lead to varying comparison scenarios between QML and CML algorithms, which might not be captured in our study. However, we chose Qiskit due to its integration with IBM Cloud Quantum Computing Services, aligning with our future plans to extend the study on real quantum machines. We anticipate that the results from the Qiskit-based study can be generalized to other QML libraries, but future work will explore additional QML algorithms to broaden the external validity by incorporating a diverse set of algorithms from multiple sources.

\textbf{Construct Validity:} One potential threat to construct validity concerns the runtime of the QSVC algorithm, which was found to be significantly longer than that of other algorithms. While this could undermine the practical applicability of QSVC in real-world scenarios.  \citet{QuantumSupremacy1} reported that in their experimental setup, an execution time of 200 seconds in a 53 qubits quantum computer is equivalent to approximately 10,000 years or \num{3.1536e11} seconds in a state-of-the-art classical supercomputer. Therefore, the observed delays in the simulated environment may not necessarily translate to the same extent in a real quantum computing environment. We anticipate that, given the rapid advancements in quantum computing technology, the QML algorithms, including QSVC, would demonstrate significantly improved efficiency in a real quantum computing environment compared to the simulated setup, mitigating the current run-time concerns.

\section{Related Work}
\label{related-work}

In machine learning, the concept of quantum computing is applied to get its unique benefit in tackling complex problems that classical computers struggle with, offering the potential for faster data processing and more efficient optimization, which can transform the field of artificial intelligence and data analysis \cite{ lloyd2014quantum, alvarez2017supervised}. Although early works in quantum computing had the goal of minimizing the complexity of any computation and making the results faster \cite{Schuld2017SimpleQ}, recent research has delved into how quantum techniques can offer distinct learning representations \cite{learningRepresentations}. 
Several instances include quantum clustering \cite{casana2020probabilistic}, quantum autoencoders \cite{lamata2018quantum}, quantum Reinforcement Learning (RL) \cite{lamata2017basic}, quantum nonlinear modeling \cite{wiebe2014quantum}, and quantum acceleration in active learning (AL) \cite{paparo2014quantum}. Quantum Software Engineering \cite{QSE2022software} is also developing at a rapid pace to make different tools and algorithms to provide software support quantum software and manage applications across diverse fields, such as finance, healthcare, and agriculture. 

Quantum-enhanced machine learning \cite{dunjko2016quantum} has the potential to advance the fields of supervised, unsupervised, and reinforcement learning, offering quadratic efficiency improvements and exponential performance gains over limited periods in various learning scenarios. The use of the classical ML approach to enhancing the performance of quantum systems is also being investigated by implementing an intelligent agent with a projective simulator \cite{tiersch2015adaptive}. This can adapt measurement directions to unknown external stray fields, providing quantum information processing devices robustness and stability. \citet{QuantumPySE} emphasize the potential of hybrid quantum-classical neural networks, error mitigation on noisy quantum devices, and the application of machine learning to enhance quantum hardware and solve physics problems.

Integration of Quantum computing into software engineering is still limited. In a study by \citet{dynamicST}, QML algorithms were examined in dynamic software testing. Our research takes a significant stride toward a similar goal, marking a pioneering effort by conducting a comprehensive analysis, comparing performance, and addressing challenges associated with QML Vs. CML algorithms. 



\section{Conclusion \& Future Work}
\label{conclusion-future}

Our investigation reveals the strengths and weaknesses of both QML and CML algorithms, by shedding light on the potential avenues for further research and improvement in applying quantum machine learning approaches for predicting software bug. This investigation provides evidence of the practical applicability of QML algorithms to detect bugs. Our investigation shows that QML algorithms are effective with a limited-size data set, which is a natural phenomenon, especially for newer applications or newer versions of an application. Our investigation shows that QML algorithms provide better results than CML algorithms in such a scenario. As QML algorithms are still being developed, more similar research needs to be performed. It can bring more effective solutions for detecting real-life bugs, even from large software systems. This investigation is an initial step that can help us take more steps forward to uncover the vast potential of quantum computing in software engineering domains.  

Researchers can take several avenues to enhance our understanding in the future studies. One promising avenue for further research could be the hybridization of QML and CML algorithms by taking benefits from both approaches. Besides hybridization, researchers can also investigate how ensembling quantum and classical algorithms, each with its unique strengths and weaknesses, can lead to a more comprehensive and accurate code defect prediction model. Ensembling can mitigate the weaknesses of individual algorithms and capitalize on their strengths, potentially resulting in higher overall performance.

\bibliographystyle{plainnat}
\footnotesize
\balance
\bibliography{citations}






\end{document}